\documentclass[conference]{IEEEtran}
\IEEEoverridecommandlockouts

\usepackage[utf8]{inputenc} 
\usepackage[T1]{fontenc}    
\usepackage{hyperref}       
\usepackage{url}            
\usepackage{booktabs}       
\usepackage{nicefrac}       
\usepackage{microtype}      
\usepackage{cite}
\usepackage{amsmath,amssymb,amsfonts}
\usepackage{algorithmic}
\usepackage{graphicx}
\usepackage{textcomp}
\usepackage{xcolor}
\usepackage{caption}

\usepackage[lofdepth,lotdepth]{subfig}  
\usepackage{multirow}                   
\usepackage{pdflscape}                  

\usepackage[colorinlistoftodos]{todonotes}

\title{Comparison of Atom Representations in Graph Neural Networks for Molecular Property Prediction\thanks{The work of A. Pocha and Ł. Maziarka was supported by the National Science Centre (Poland) grant no. 2019/35/N/ST6/02125. The work of T. Danel was supported by the National Science Centre (Poland) grant no. 2020/37/N/ST6/02728.}}

\author{%
  \IEEEauthorblockN{%
    \parbox{\linewidth}{\centering
      Agnieszka Pocha\IEEEauthorrefmark{1},
      Tomasz Danel\IEEEauthorrefmark{1}\IEEEauthorrefmark{2},
      Sabina Podlewska\IEEEauthorrefmark{3},
      Jacek Tabor\IEEEauthorrefmark{1} and
      Łukasz Maziarka\IEEEauthorrefmark{1}\IEEEauthorrefmark{2}%
    }%
  }%
  \IEEEauthorblockA{%
    \IEEEauthorrefmark{1}Faculty of Mathematics and Computer Science, Jagiellonian University, Kraków, Poland \\
    \IEEEauthorrefmark{2}Ardigen, Kraków, Poland \\
    \IEEEauthorrefmark{3}Maj Institute of Pharmacology, Polish Academy of Sciences, Kraków, Poland \\
    Email: agnieszka.pocha@doctoral.uj.edu.pl, lukasz.maziarka@ii.uj.edu.pl%
  }%
}

\begin{document}

\maketitle

\begin{abstract}
  Graph neural networks have recently become a standard method for analysing chemical compounds. In the field of molecular property prediction, the emphasis is now put on designing new model architectures, and the importance of atom featurisation is oftentimes belittled. When contrasting two graph neural networks, the use of different atom features possibly leads to the incorrect attribution of the results to the network architecture. To provide a better understanding of this issue, we compare multiple atom representations for graph models and evaluate them on the prediction of free energy, solubility, and metabolic stability. To the best of our knowledge, this is the first methodological study that focuses on the relevance of atom representation to the predictive performance of graph neural networks.
\end{abstract}


\section{Introduction}

Graph convolutional neural networks (GCNs) are state-of-the-art models for predicting molecular properties. As the input, they use molecular graphs in which vertices represent atoms and edges represent the chemical bonds.
Since graph-based models were shown to outperform models based on molecular fingerprints~\cite{duvenaud2015convolutional}, the interest in GCNs increased, which resulted in proposing new models for molecular property prediction~\cite{coley2017convolutional,gilmer2017neural,schutt2018schnet,yang2019analyzing,klicpera2020directional,maziarka2020molecule,danel2020spatial,rong2020self,song2020communicative}.

Since the beginning, the main focus of the deep learning community has been placed on developing better machinery for processing the graph data. For instance \cite{velivckovic2017graph} introduce the attention mechanism for GCNs and \cite{li2017learning} introduce a dummy super node -- an artificial node connected to all nodes in the graph which is responsible for learning the graph-level representation. At the same time, authors of new methods oftentimes neglect the impact of the used atomic representation.
Therefore, atoms are represented in a different way for each new graph-based model, which may lead to the unfair attribution of the achieved results solely to the developed processing methods. 

There is a need for systematic comparison of graph representations which seems independent from the choice of architecture. In this work, we compare different atomic representations using simple GCNs~\cite{kipf2016semi} and evaluate them on multiple datasets.

Initially, we study the influence of single atomic features in isolation and show that some of them encode more useful information than others. Subsequently, we examine few exemplary representations used in literature and show their differences in terms of accuracy and generalisation gap.

Furthermore, we analyse the results qualitatively by showing molecules which are most difficult to predict for GCNs with a given representation. Finally, we use t-SNE~\cite{van2008visualizing} to show that molecules with the highest prediction error tend to cluster together.

\section{Related work}

Two main components of molecular property prediction are the representation of chemical compounds and the algorithm used to calculate the property values. The classical machine learning methods which were used to find the relationship between chemical structures of molecules and their properties used simple 1D molecular descriptors, e.g. lipophilicity or electron density, which were plugged into machine learning models to create predictions of more complex molecular properties~\cite{kubinyi1997qsar}. Shortly afterwards, these descriptors were replaced by features derived from the structure of molecules. 

A prominent example of structural compound descriptors are molecular fingerprints, which are typically a mapping from chemical substructures to numeric feature vectors. The vectors constructed in this way can become an input to machine learning models, e.g. random forests, support vector machines, or neural networks, in order to find quantitative structure-property relationships (QSPR).

ECFP fingerprint~\cite{rogers2010extended} is one of the most commonly used fingerprints in this setup~\cite{perryman2016predicting, laufkotter2019combining}. To calculate this molecular representation, the algorithm uses a hash function that encodes fragments contained in the molecule. The crucial part of this encoding is the atomic representation that makes the individual atoms in fragments distinguishable. As their representation, Rogers and Hahn~\cite{rogers2010extended} use the number of non-hydrogen neighbours, the valence minus the number of hydrogens, the atomic number, the atomic mass, the atomic charge, the number of attached hydrogens, and inclusion in rings.

With the development of recurrent neural networks for natural language processing SMILES~\cite{doi:10.1021/ci00057a005}, a string representation of a molecular graph, became a frequent choice for both molecular property prediction~\cite{jastrzkebski2016learning} and molecule generation~\cite{kusner2017grammar, olivecrona2017molecular, popova2018deep, gomez2018automatic}.

Currently, the neural representations of molecules are displacing molecular fingerprints as graph neural networks can learn a molecular representation that is tailored to the prediction task. The problem of the selection of atomic representation remains relevant also for these state-of-the-art methods as they require atom features and molecular graph topology at the input. The atomic representations diverge, starting from the first works on GCNs. For example, Kearnes \textit{et al.}~\cite{kearnes2016molecular} use atom types, chirality, formal and partial charge, ring sizes, hybridization, hydrogen bonding, and aromaticity. Gilmer \textit{et al.~}\cite{gilmer2017neural} use one-hot encoding of the atom type alone, whereas Coley \textit{et al.}~\cite{coley2017convolutional} only encode 10 most common atom types along with the number of atom heavy neighbours, the number of hydrogen neighbours, aromaticity, formal charge and inclusion in a ring. Liu \textit{et al.}~\cite{liu2019chemi} expand the one-hot representation to 23 most common atom types and add information about vdW and covalent radius of the atom. However, they do not use information about atom neighbourhood. 
Yang \textit{et al.}~\cite{yang2019analyzing} extend one-hot encoding to 100 dimensions and add information about atom's chirality, atomic mass, hybridization and number of bonds the atom is involved in.
Moreover, some of the models additionally use bond vector representations. 

A huge diversity of atomic representations makes it difficult to compare performance between different models. One must take into consideration that the differences in performance might arise not only from the choices concerning the architecture, but also the representation being used. To the best of our knowledge, there does not exist any extensive study of atomic representations in graph neural networks, so currently the choice of the used atomic features is subjective. In this study, we indicate the importance of the atomic features and measure their impact on the graph-based models performance.

\section{Data and Methods}

In this section, we first describe the representations used in our experiments and follow with the details of the GCN architecture and the model selection method. Next, we briefly summarise the statistical methods used to anylyse the results and finally, we describe the datasets chosen for evaluation.

\subsection{Atom representations}
We represent a molecule with $N$ atoms as an undirected graph $\mathcal{G}=(X, A)$, where $X\in\mathbb{R}^{N\times D}$ is the atomic representation matrix, $A\in\mathbb{R}^{N\times N}$ is the graph adjacency matrix, and $D$ is the number of atomic features. We chose five commonly used atom features: the atom type (one-hot encoded atom symbol), the number of heavy (non-hydrogen) atom neighbours, the number of attached hydrogens, formal charge, inclusion in a ring, and aromaticity. We consider 4 representation groups: 
\begin{itemize}
    \item using all the atomic features,
    \item using only one-hot encoded atom types,
    \item using exactly one atomic feature besides the atom type,
    \item using all atomic features but one.
\end{itemize}
The details are given in Table~\ref{tab:representations}.

\addtolength{\tabcolsep}{-1.5pt} 
\begin{table}[!htb]
    \caption{Features included in each of the 12 atom representations.}
    \centering
    \begin{tabular}{ccccccccccccc}
        \toprule
        & 1 & 2 & 3 & 4 & 5 & 6 & 7 & 8 & 9 & 10 & 11 & 12 \\
        \midrule
        atom type     & \checkmark & \checkmark & \checkmark & \checkmark & \checkmark & \checkmark & \checkmark & \checkmark
        & \checkmark & \checkmark & \checkmark & \checkmark\\
        neighbors     & \checkmark & & \checkmark & & & & &
        & \checkmark & \checkmark & \checkmark & \checkmark\\
        hydrogens     & \checkmark & & & \checkmark & & & &
        \checkmark & & \checkmark & \checkmark & \checkmark\\
        formal charge & \checkmark & & & & \checkmark & & &
        \checkmark & \checkmark & & \checkmark & \checkmark\\
        in ring       & \checkmark & & & & & \checkmark & &
        \checkmark & \checkmark & \checkmark & & \checkmark\\
        aromatic      & \checkmark & & & & & & \checkmark &
        \checkmark & \checkmark & \checkmark & \checkmark &\\
        \bottomrule
    \end{tabular}
    \label{tab:representations}
\end{table}
\addtolength{\tabcolsep}{1.5pt}

\subsection{Model}

We use graph neural network implementation based on~\cite{kipf2016semi}. Namely, we use the following graph convolution formula:
\begin{equation}
    H^{(l+1)} = D^{-\frac{1}{2}}\hat{A}D^{-\frac{1}{2}}H^{(l)}W^{(l)},
\end{equation}
where $\hat{A}=A+I$ is the graph adjacency matrix including self-loops, $D_{ii}=\sum_j \hat{A}_{ij}$, $H^{(l)}$ is the node representation matrix in the $l$-th layer, and $W^{(l)}$ is a trainable weight matrix. The node representation at the input to the first layer is the atomic representation matrix ($H^{(0)} = X$).

\paragraph{Model selection} The best performing architectures were found using random search. All neural networks consist of graph convolutional layers followed by dense layers, and vary by: number of convolutional layers, number of channels in each convolutional layer, number of dense layers, size of dense layers, dropout, batchnorm, learning rate, batch size, and learning rate scheduler. The number of channels in convolutional layers and the size of hidden layers are equal in all models. A detailed description of the hyperparameter space can be found in Table~\ref{tab:grid}. All models were trained for 750 epochs using Adam and MSE loss. 

\begin{table}[htb!]
\centering
\caption{Hyperparameters considered in our experiments}
\begin{tabular}{@{}ll@{}}
\toprule
hyperparameter                              & values considered                 \\
\midrule
number of conv. layers             & 1, 3, 5                            \\
number of channels in conv. layers & 16, 64, 256                        \\
number of dense layers                     & 1, 3                               \\
size of dense layers                       & 16, 64, 256                        \\
dropout                                    & 0.0, 0.2                           \\
batchnorm                                  & True, False                        \\
batchsize                                  & 8, 32, 128                         \\
learning rate                              & .01, .001, .0001, .00001, .000001  \\
scheduler                                  &\begin{tabular}{@{}l@{}}no scheduler, \\ decrease after 50\% of epochs, \\ decrease after 80\% of epochs\end{tabular} \\
\bottomrule
\end{tabular}\label{tab:grid}
\end{table}

We use the same set of 100 randomly sampled hyperparameter configurations for all the datasets. Each architecture was trained three times to accommodate for variance resulting from random initialisation.

\subsection{Statistical methods} 

To compare atom features, we picked the best architecture found in random search for each representation. We performed one- and two-tailed Wilcoxon tests with Bonferroni correction to analyze the differences between representations.

\begin{figure*}
    \centering
    \includegraphics[width=\textwidth]{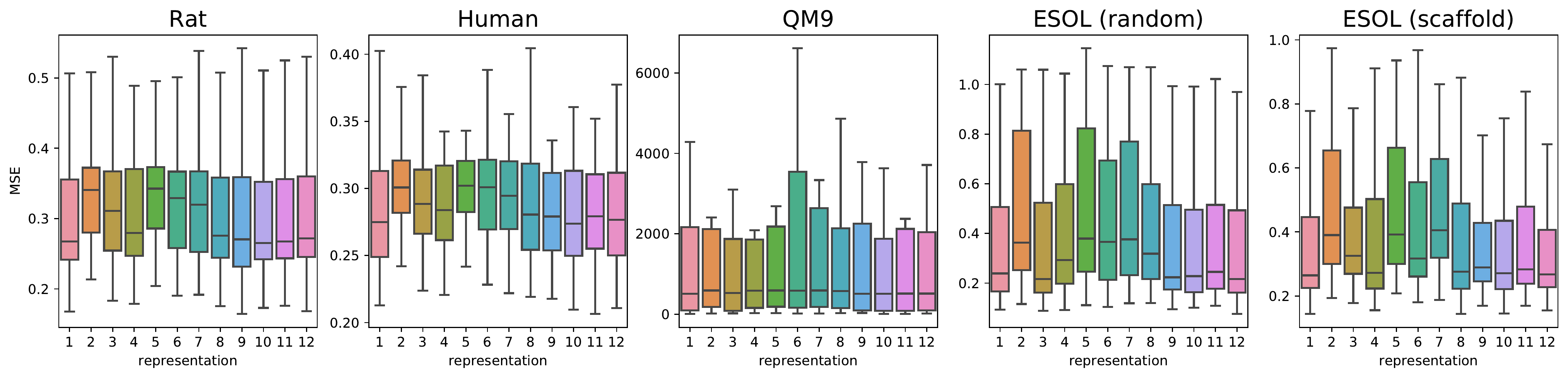}
    \caption{Distribution of mean square error on the test set of all models trained with the selected representation.}
    \label{fig:boxplots}
\end{figure*}

\begin{table*}[h!]
    \centering
    \caption{Average test mean squared error of all models trained with different representations (no model selection).}
    \begin{tabular}{c | ccccc}
    \toprule
    representation &                rat &              human &         qm9-random &      esol-random &    esol-scaffold \\
    \midrule
    1 &  0.301 $\pm$ 0.102 &  0.291 $\pm$ 0.090 &  34100 $\pm$ 65370 &  0.39 $\pm$ 0.29 &  0.37 $\pm$ 0.21 \\
    2 &  0.339 $\pm$ 0.087 &  0.314 $\pm$ 0.075 &  35990 $\pm$ 65780 &  0.49 $\pm$ 0.29 &  0.47 $\pm$ 0.21 \\
    3 &  0.322 $\pm$ 0.101 &  0.303 $\pm$ 0.085 &  34820 $\pm$ 65340 &  \textbf{0.37 $\pm$ 0.29} &  0.41 $\pm$ 0.19 \\
    4 &  0.311 $\pm$ 0.104 &  0.299 $\pm$ 0.073 &  35140 $\pm$ 65300 &  0.42 $\pm$ 0.29 &  0.38 $\pm$ 0.22 \\
    5 &  0.342 $\pm$ 0.092 &  0.315 $\pm$ 0.078 &  35740 $\pm$ 65670 &  0.49 $\pm$ 0.30 &   0.47 $\pm$ 0.20 \\
    6 &  0.329 $\pm$ 0.101 &  0.310 $\pm$ 0.083 &  35210 $\pm$ 65410 &  0.46 $\pm$ 0.29 &  0.42 $\pm$ 0.21 \\
    7 &  0.326 $\pm$ 0.100 &  0.308 $\pm$ 0.084 &  35930 $\pm$ 65640 &  0.48 $\pm$ 0.28 &  0.47 $\pm$ 0.19 \\
    8 &  0.303 $\pm$ 0.098 &  0.295 $\pm$ 0.083 &  34460 $\pm$ 65300 &  0.44 $\pm$ 0.28 &  0.37 $\pm$ 0.22 \\
    9 &  0.299 $\pm$ 0.104 &  0.295 $\pm$ 0.083 &  34670 $\pm$ 65290 &  0.38 $\pm$ 0.28 &   0.38 $\pm$ 0.20 \\
    10 &  \textbf{0.297 $\pm$ 0.093} &  \textbf{0.289 $\pm$ 0.078} &  \textbf{34040 $\pm$ 65270} &  0.38 $\pm$ 0.29 &  \textbf{0.36 $\pm$ 0.21} \\
    11 &  0.300 $\pm$ 0.097 &  0.294 $\pm$ 0.084 &  34250 $\pm$ 65360 &  0.39 $\pm$ 0.28 &   0.37 $\pm$ 0.20 \\
    12 &  0.302 $\pm$ 0.099 &  0.290 $\pm$ 0.080 &  34060 $\pm$ 65340 &  0.38 $\pm$ 0.29 &   \textbf{0.36 $\pm$ 0.20} \\
    \bottomrule
    \end{tabular}
    \label{tab:boxplot-table}
\end{table*}

\subsection{Datasets}

For evaluation we chose four datasets that represent a wide range of molecular property prediction tasks. For the ESOL dataset, we use two different methods of splitting the data, random split and scaffold split~\cite{bemis1996properties}, to examine if the choice of the splitting method affects the performance of models trained with different representations. The datasets used in our experiments are:

\paragraph{QM9} a dataset for predicting quantum properties~\cite{ruddigkeit2012enumeration}. We randomly sampled 5K molecules for training, 1k molecules for validation, and 10\% of the dataset (13K molecules) for the test set. The models were trained to predict g298 (Free energy at 298.15 K (unit: Hartree)).

\paragraph{ESOL} a water solubility prediction dataset of 1128 samples~\cite{delaney2004esol}. We report results on both random split from \cite{maziarka2020molecule} and 80-10-10 scaffold split.

\paragraph{HUMAN and RAT} datasets for metabolic stability prediction from~\cite{podlewska2018metstabon}. Only records with the source being 'Liver', 'Liver microsome', or 'Liver microsomes' were used, resulting in 3578 and 1819 samples, respectively. In case of multiple measurements for the same molecule, the median of the measurements is used. The stability values are expressed in hours and log scaled. 10\% of data was left out for testing and the remaining samples were divided into 5 cross-validation folds using random stratified split.

\section{Results}

\subsection{Quantitative analysis}

In Figure~\ref{fig:boxplots}, we compare the performance of models trained with different representations. Datasets and representations are on the x-axis and on the y-axis the distribution of mean square error on  the test set of all models trained with the selected representation.

\begin{figure}[htb!] 
    \centering
    \includegraphics[width=0.85\linewidth]{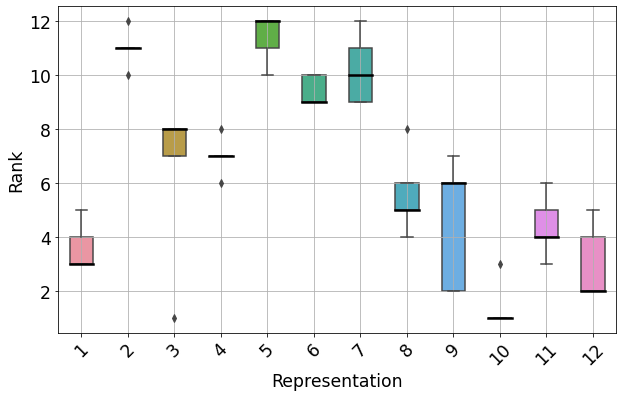}
    \caption{Rankings obtained for every given representation on all datasets. The median ranking is marked with a bold line.}
    \label{fig:rankplot}
\end{figure}

In Table~\ref{tab:boxplot-table}, one can see detailed results and Figure~\ref{fig:rankplot} presents the box plot with rankings obtained by the representations on all datasets. 
The best scores are obtained by models trained with representation $10$ (no formal charge), which beat models trained with other representations in 4 out of 5 tasks.

\begin{figure*}
    \centering
    \includegraphics[width=\textwidth]{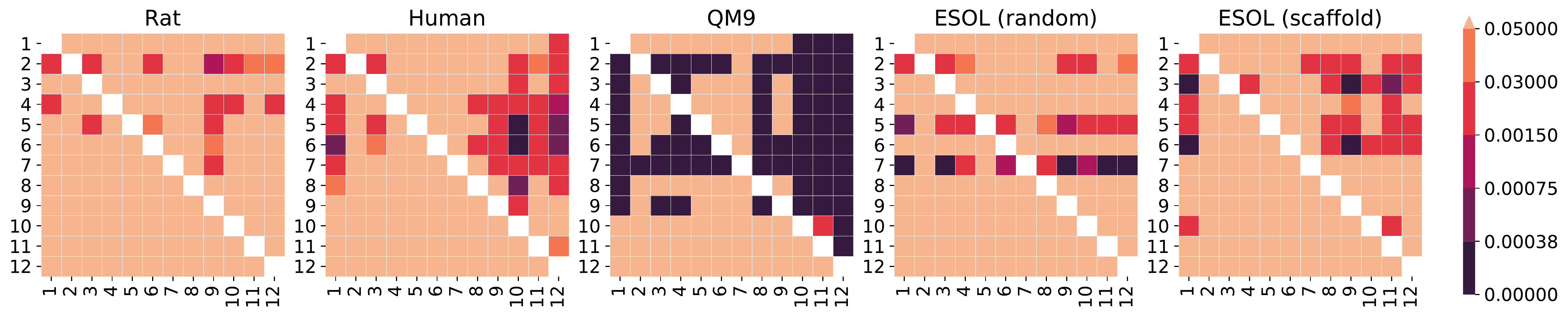}
    \caption{P-values of one-tailed Wilcoxon tests between the best models trained on each representation. The value in $i$-th row and $j$-th column corresponds to the alternative hypothesis saying that the median squared error of $i$-th representation is greater than the median of $j$-th representation (superior representations have darker columns, and inferior ones have darker rows). The darkest cells are statistically significant with Bonferroni correction.}
    \label{fig:wilcoxon}
\end{figure*}

\begin{figure*}[h!]
\centering
\subfloat[Subfigure 1][overall]{
\includegraphics[width=0.24\textwidth]{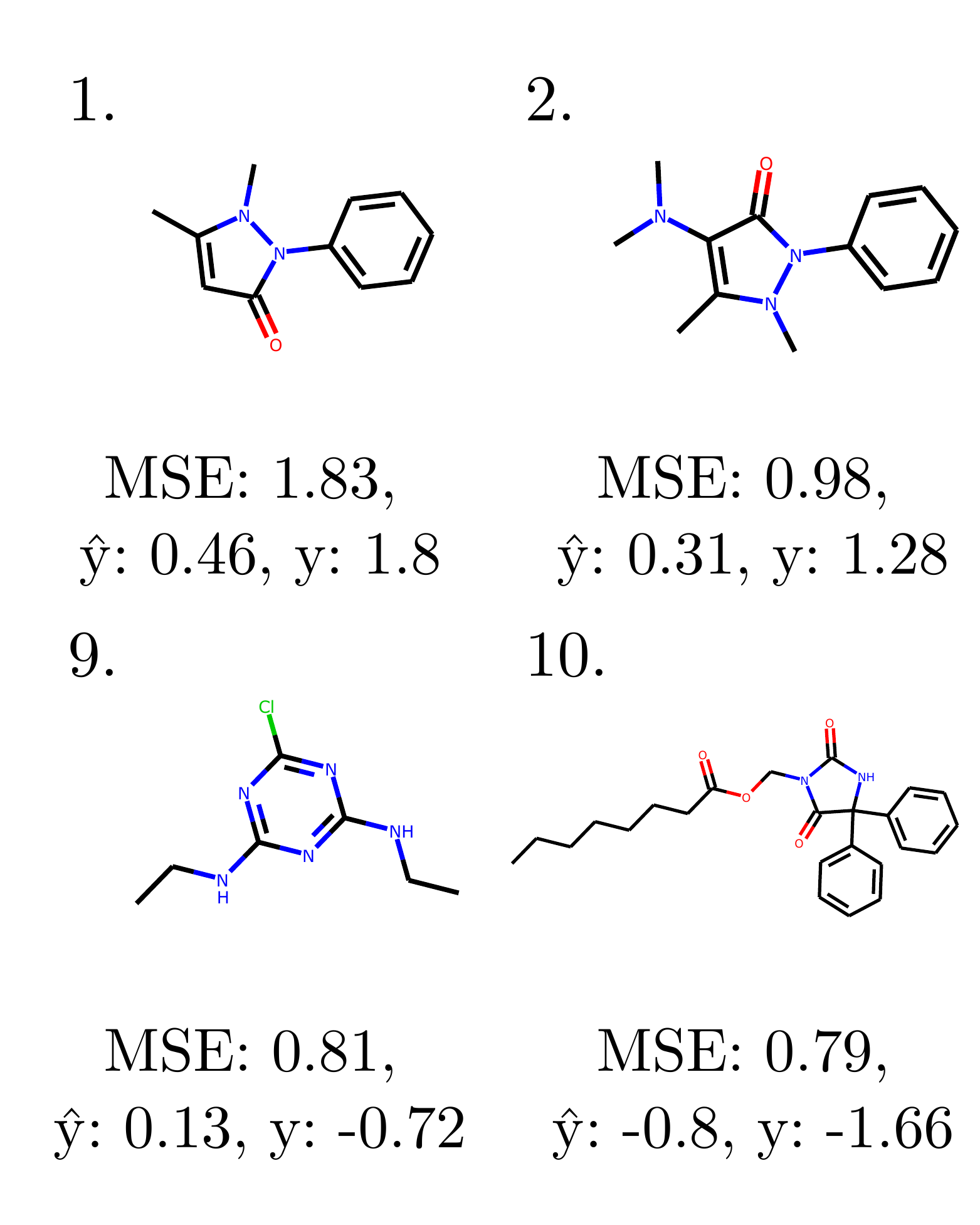}
\label{subfig:weirdos-esol-scaffold-overall}}
\subfloat[Subfigure 2][repr-2 (empty)]{
\includegraphics[width=0.24\textwidth]{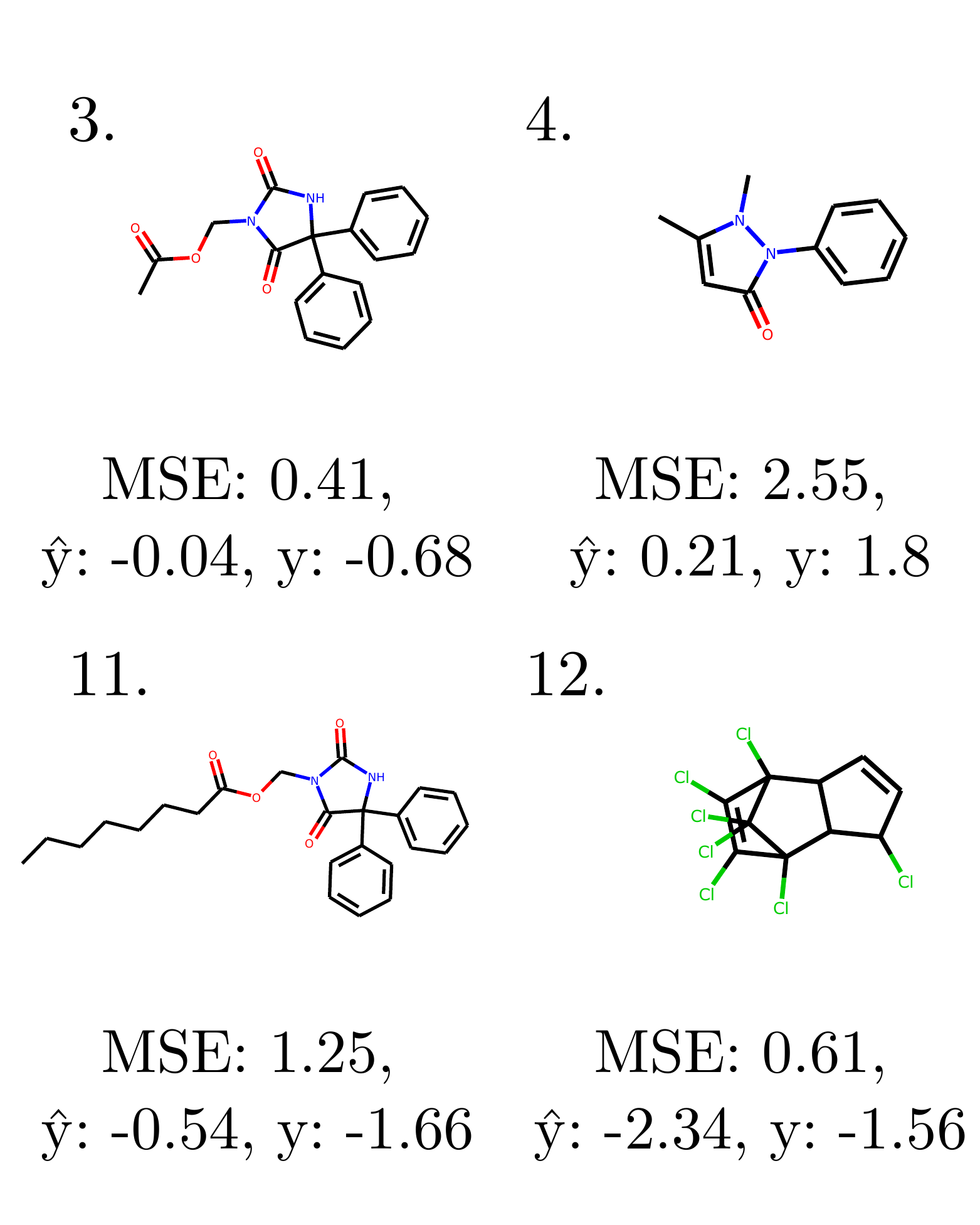}
\label{subfig:weirdos-esol-scaffold-2}}
\subfloat[Subfigure 3][repr-11 (no rings)]{
\includegraphics[width=0.24\textwidth]{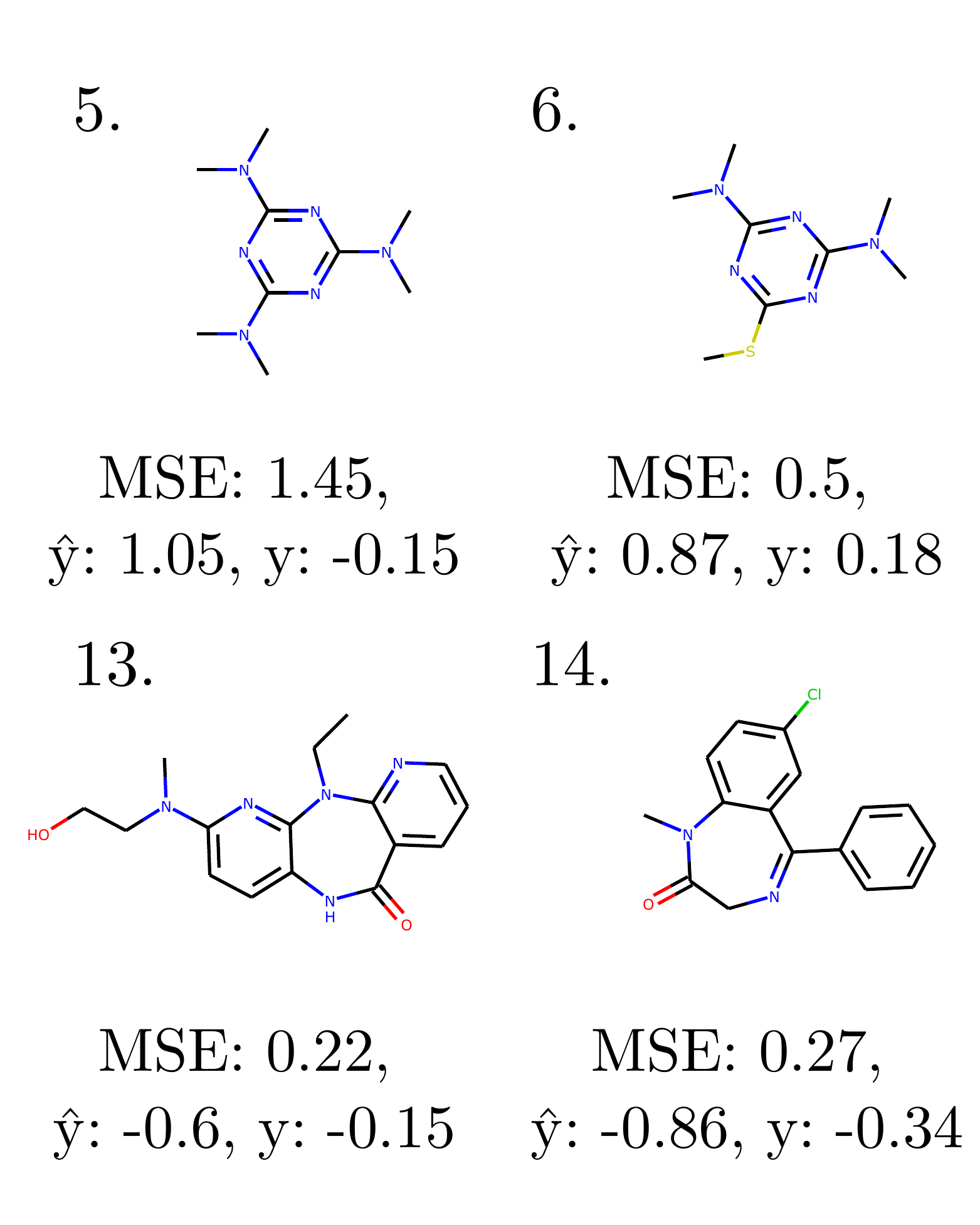}
\label{subfig:weirdos-esol-scaffold-11}}
\subfloat[Subfigure 4][repr-12 (no aromatic)]{
\includegraphics[width=0.24\textwidth]{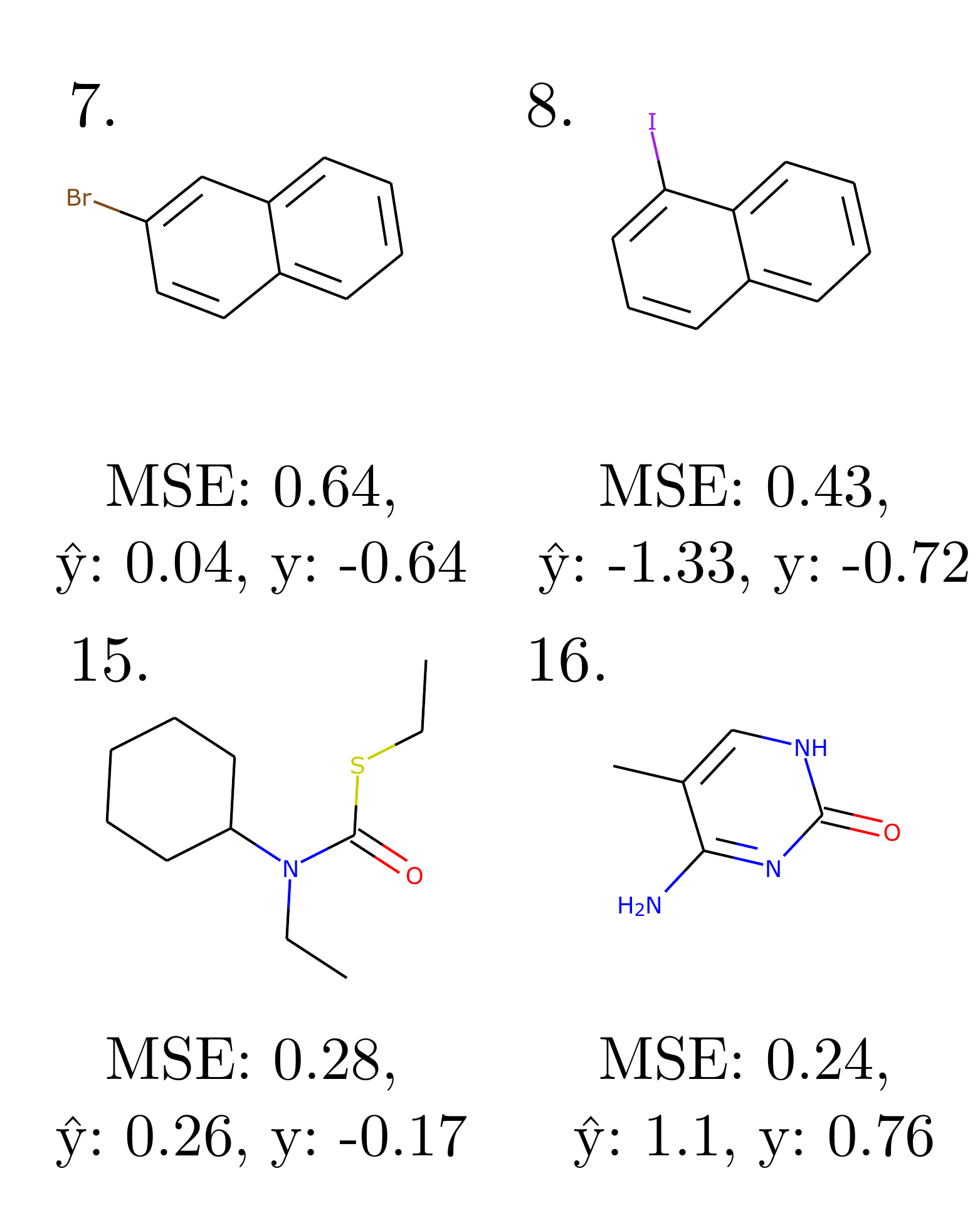}
\label{subfig:weirdos-esol-scaffold-12}}
\caption{The worst predicted molecules in the ESOL (scaffold) dataset. Plots show compounds with the highest MSE in all representations (a), and MSE higher than in other representations (b-d); $\hat{y}$ is the average predicted value, and $y$ is the true value (standardised).}
\label{fig:weirdos-esol-scaffold}
\end{figure*}

In order to systematically study the error distributions, we ran Wilcoxon tests for pairwise representation comparisons. The \mbox{p-values} of one-sided tests are plotted in Figure~\ref{fig:wilcoxon}. We observe that many representations are equivalent even before applying the Bonferroni correction (p $\geq0.05$), e.g. in RAT the lowest p-value is above the level of significance (p $\geq 0.002$ in a two-tailed Wilcoxon test, while the significant differences should be below 0.05 / 66 pairwise tests). The differences between representations are most apparent in QM9, which is the biggest dataset in the comparison (p $\leq \nicefrac{0.05}{66}$ in all two-tailed Wilcoxon tests besides the ones between representations 3-5, 5-9, and 10-11).

There are several patterns that can be noted in the heatmaps.

\begin{enumerate}
\item Atom representations with almost full set of features are usually comparable with each other (bright area in the bottom left corner) and better than nearly empty feature vectors (dark area in the top right corner).
\item There are features that perform significantly worse than others when used alone, e.g. including only aromaticity (repr.~7) yields almost as poor results as using no atom features in QM9 and ESOL with a random split. On the other hand, adding information about heavy neighbors and hydrogens (repr. 3 and 4) gives the biggest performance boost across all datasets.
\item Removing features related to aromaticity (repr. 12), inclusion in a ring (repr. 11), and formal charges (repr. 10) can improve model quality, compared with the full representation (repr. 1).
\end{enumerate}

\subsection{Qualitative analysis} 

\begin{figure*}[h!]
    \centering
    \subfloat[Subfigure 1][overall]{
    \includegraphics[width=0.24\textwidth]{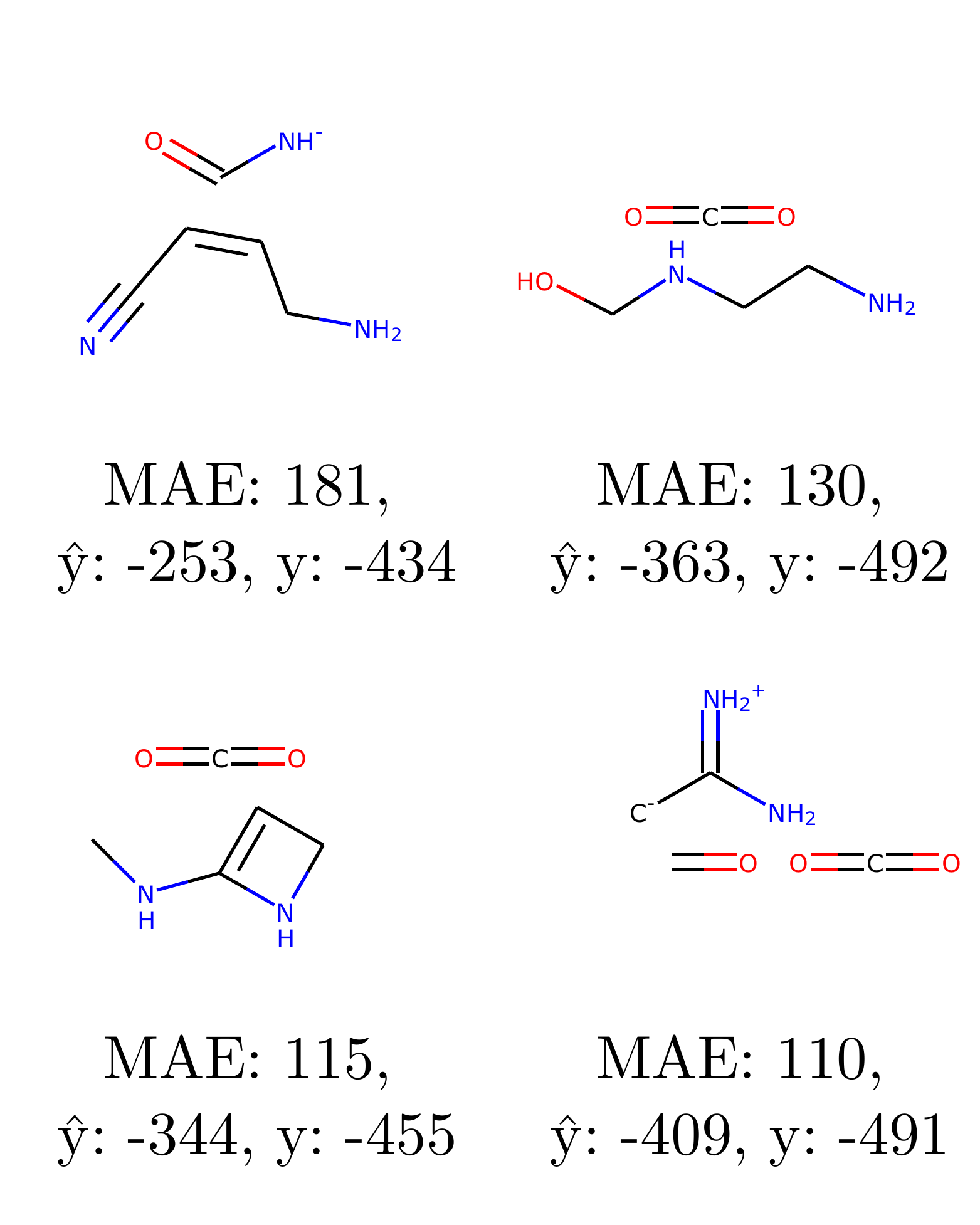}
    \label{subfig:weirdos-qm9-overall}}
    \subfloat[Subfigure 2][repr-4 (hydrogens)]{
    \includegraphics[width=0.24\textwidth]{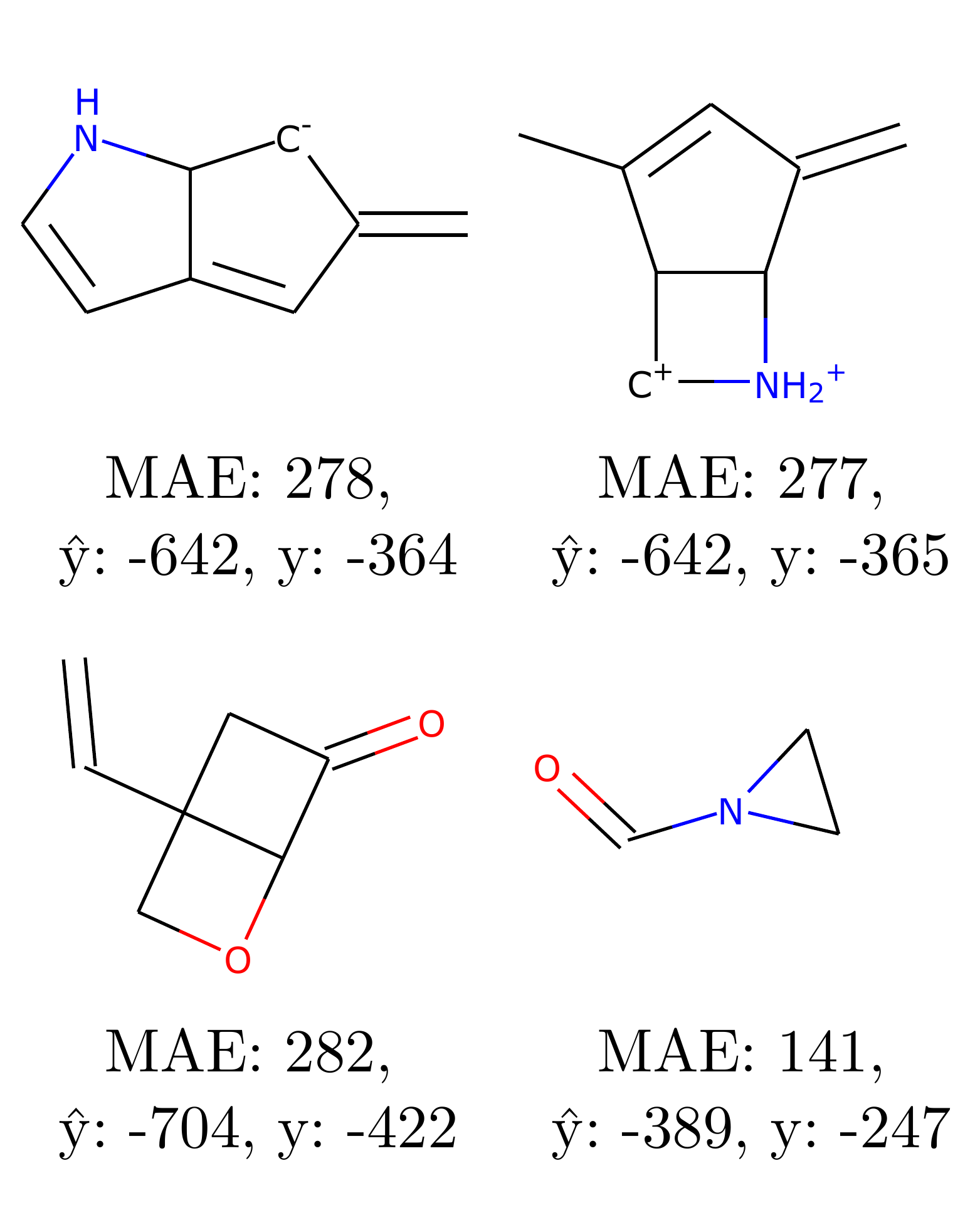}
    \label{subfig:weirdos-qm9-4}}
    \subfloat[Subfigure 3][repr-8 (no neighbors)]{
    \includegraphics[width=0.24\textwidth]{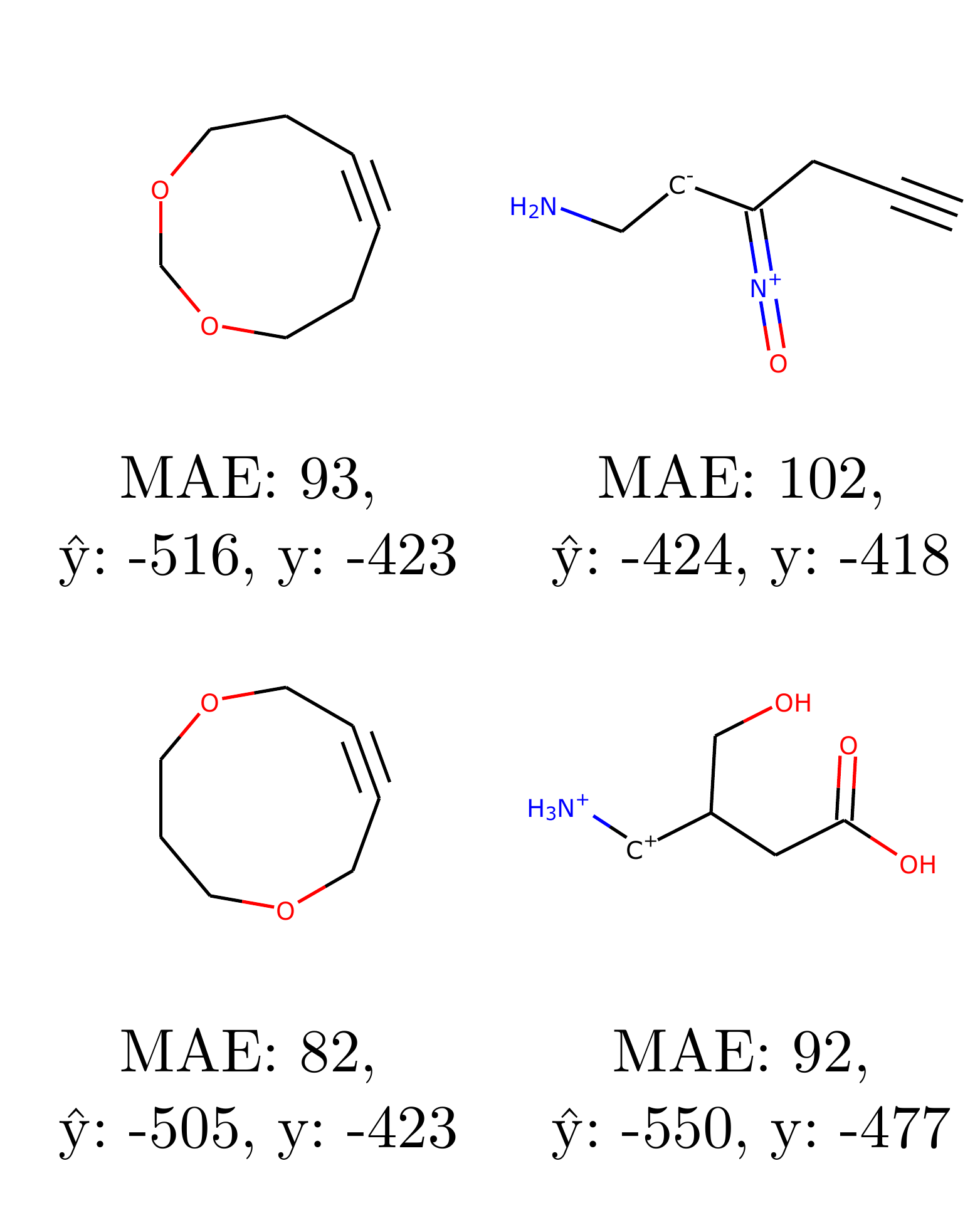}
    \label{subfig:weirdos-qm9-8}}
    \subfloat[Subfigure 4][repr-9 (no hydrogens)]{
    \includegraphics[width=0.24\textwidth]{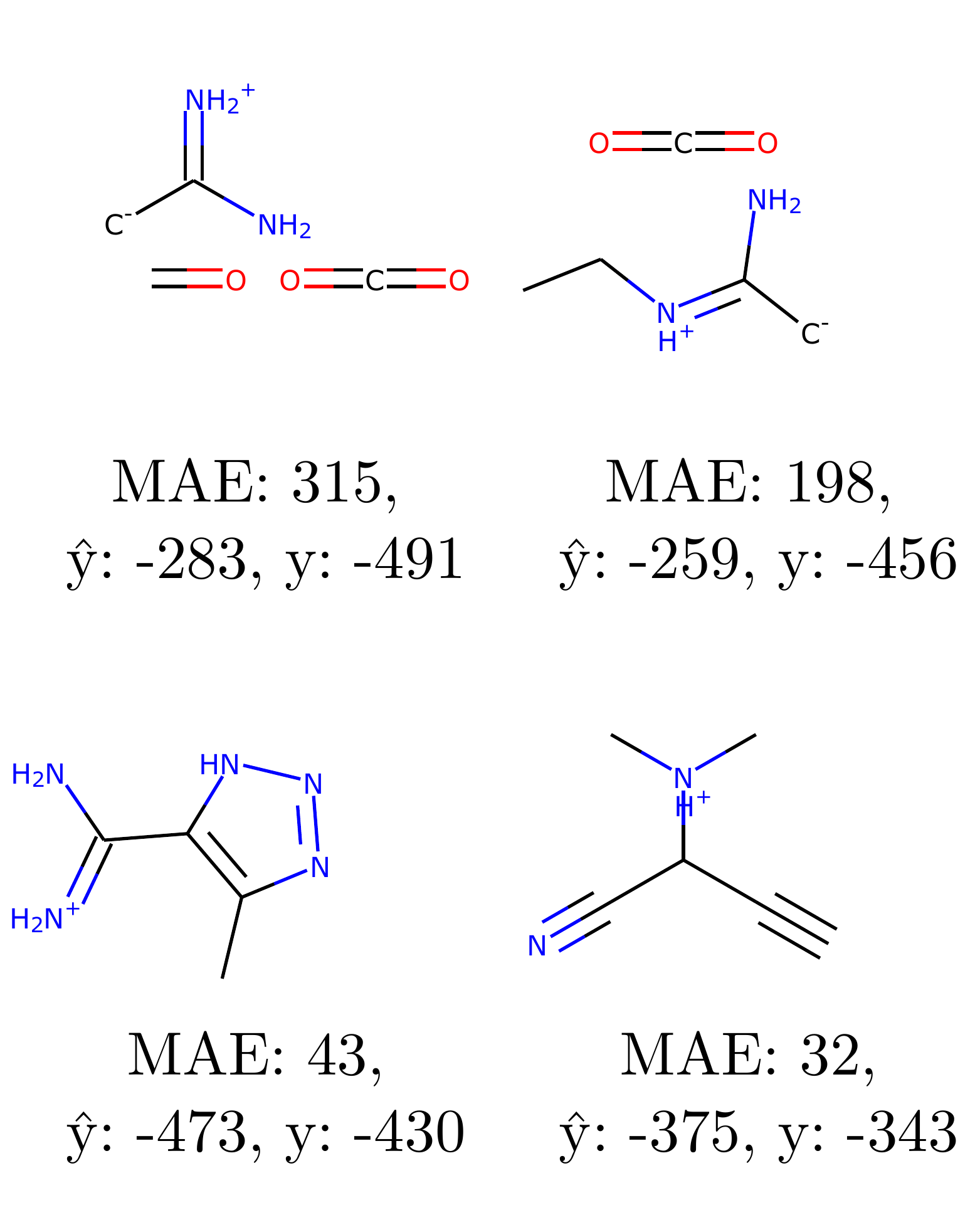}
    \label{subfig:weirdos-qm9-9}}
    \caption{The worst predicted molecules in the QM9 dataset. Plots show compounds with the highest MAE in all representations (a) and MAE higher than in other representations (b-d); $\hat{y}$ is the average predicted value, and $y$ is the true value (standardized).}
    \label{fig:weirdos-qm9}
\end{figure*}

Figure~\ref{fig:weirdos-esol-scaffold} shows molecules with the highest mean errors of solubility prediction for all representations jointly and for three selected ones. To pick molecules that are predicted worse by a given representation, we calculated a margin between the mean error of this representation and the highest mean error of the remaining representations. To put it more precisely, for each compound we calculate predictions using the best model for each representation $\hat{y}_1,\dots,\hat{y}_{12}$ and compare these predictions with the true label $y$. Next, we sort the compounds by the following value:
\begin{equation}
    m_i = \max\left(0, \varepsilon(y - \hat{y}_i) - \max_{\substack{j=1,\dots,12 \\ i\neq j}}\varepsilon(y-\hat{y}_j)\right),
\end{equation}
where $\varepsilon:\mathbb{R}\to\mathbb{R_+}$ is an error function (e.g. MSE or MAE), and $m_i$ is the error margin of the compound for the $i$-th representation.

We observe that using only the topological graph information and no atom features besides the atom type produces similar structures to those that are on average worst predicted by all representations. For instance, the molecule with a long aliphatic chain (molecule~11) is predicted as more soluble probably because the model with no atom features cannot differentiate between saturated and unsaturated chains. Similarly, the compound with a cyclohexane ring (molecule~15) could be predicted as more soluble due to the lack of aromaticity information -- the aromatic counterpart of the cyclohexane, a benzene, is more soluble in water. Also, we note that the representation without information about ring inclusion often makes mistakes for compounds with non-aromatic rings or nitrogens in rings.

Similar results for QM9 dataset can be found in Figure~\ref{fig:weirdos-qm9}. In these selected representations we again observe recurring patterns. For example, representation 9, which does not contain information about attached hydrogens, poorly predicts compounds with nitrogen cations. Similarly, representation 8, which misses the information about the number of heavy neighbours, obtains the highest error values for branched structures with carbocations or carbanions.

\begin{figure}
    \centering
    \includegraphics[width=0.95\linewidth]{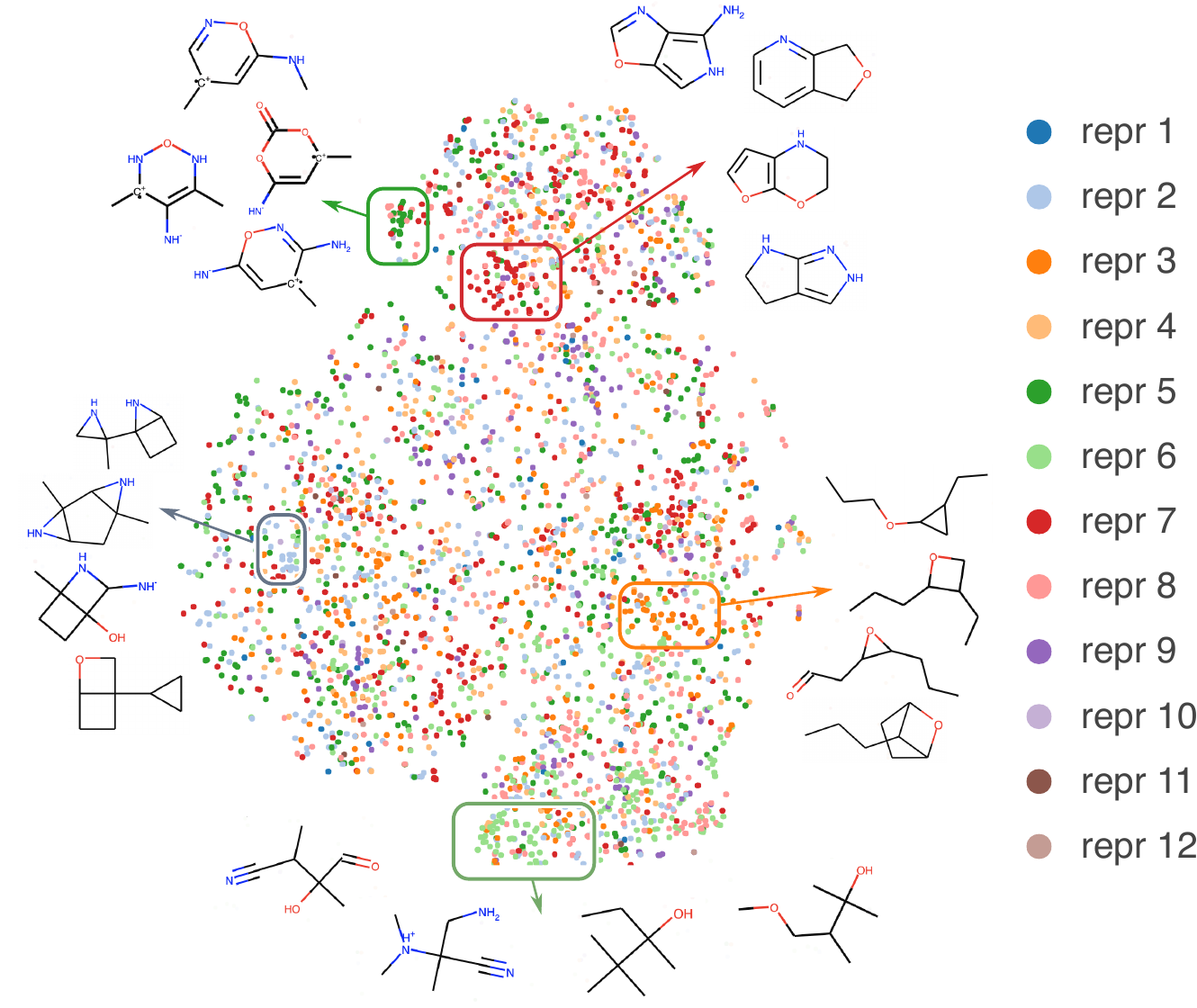}
    \caption{t-SNE map of QM9 compounds coloured by the representation with the highest prediction error (MAE). The algorithm uses the ECFP fingerprints and Tanimoto distance.}
    \label{fig:tsne}
\end{figure}

To confirm that some representations are more prone to committing errors where certain patterns appear in the molecular structure, in Figure~\ref{fig:tsne} we plotted a t-SNE map of QM9 compounds. For each representation, we selected at most 500 compounds with the highest error margin over other representations in the testing set. Additionally, the error margin was averaged over top 5 models for each representation to make the resulting map less dependent on the single training runs. Next, we calculated ECFP fingerprints to encode chemical structure of the molecules. This representation was used as an input to the t-SNE algorithm~\cite{van2008visualizing} with Tanimoto metric used for calculating distances in the fingerprint space. In the plot, each colour corresponds to the representation with the highest prediction error. To find compound clusters, we used the DBSCAN clustering algorithm~\cite{ester1996density} with $\epsilon=4$. In Figure~\ref{fig:tsne}, we plotted 5 out of 27 found clusters along with 4 compounds sampled from each of them.

We observe that small clusters of one colour form in the t-SNE map. These clusters correspond to structural motifs that confuse the model which was trained with the given representation. This observation suggests that some representations fail to predict certain structural patterns due to inductive biases. For example, representation 6 (only inclusion in rings) tends to make errors for the structures with many branches and no rings (depicted at the bottom of Figure~\ref{fig:tsne}). Interestingly, representation 5 incorrectly predicts a group of compounds with a carbocation in a ring even though it contains information about formal charge. A similar cluster corresponds to representation 7 (only aromaticity) that makes mistakes predicting fused bicyclic compounds which are partly aromatic.

\subsection{Literature representations}
The goal of this section is to put our analysis in a wider context by comparing different representations present in literature when used with the same architecture. To this end we repeat our experiments with additional representations, namely:

\begin{enumerate}
    \item[L1)] 33 dimensional representation from~\cite{liu2019chemi}.
    \item[L2)] 38 dimensional representation from~\cite{li2019deepchemstable}.
    \item[L3)] 58 dimensional representation from~\cite{duvenaud2015convolutional}.
    \item[L4)] 127 dimensional representation from~\cite{yang2019analyzing}.
\end{enumerate}

We also note that representation 1 (full) from the previous section was used in~\cite{coley2017convolutional}. The details of all representations are given in Table~\ref{tab:atom_emb_lit}. All of these representations include information whether an atom is in an aromatic system, and the type of the atom -- though the number of encoded atomic types ranges from $12$ in representations 1 and L2 up to $100$ in representation L4. Moreover, most of the representations include information about atom's (heavy) neighbours and formal charge.

\begin{table}[htb!]
    \centering
    \caption{Different featurisation methods used in the literature.}
    \begin{tabular}{ c c }
        \toprule
        \multicolumn{2}{c}{\textbf{representation 1~\cite{coley2017convolutional}}} \\
        size & description \\
        \midrule
        $12$ & one-hot vector specifying the type of atom  \\
        $6$ & number of heavy neighbours as one-hot vector \\
        $5$ & number of hydrogen atoms as one-hot vector  \\
        $1$ & formal charge \\
        $1$ & is in a ring  \\
        $1$ & is in aromatic system  \\
        \midrule \midrule
        \multicolumn{2}{c}{\textbf{representation L1~\cite{liu2019chemi}}} \\
        size & description \\
        \midrule
        $23$ & one-hot vector specifying the type of atom  \\
        $2$ & vdW radius and covalent radius of the atom \\
        $6$ & \begin{tabular}{l}\shortstack{for each size of ring (3-8) \\ the number of rings that include this atom}\end{tabular}  \\
        $1$ & is in aromatic system \\
        $1$ & electrostatic charge of this atom  \\
        \midrule \midrule
        \multicolumn{2}{c}{\textbf{representation L2~\cite{li2019deepchemstable}}} \\
        size & description \\
        \midrule
        $12$ & one-hot vector specifying the type of atom  \\
        $6$ & number of heavy neighbours as one-hot vector \\
        $5$ & number of hydrogen atoms as one-hot vector  \\
        $6$ & implicit valence as one-hot vector \\
        $1$ & is in aromatic system  \\
        $1$ & number of radical electrons  \\
        $5$ & hybridisation type as one-hot vector \\
        $1$ & formal charge \\
        $1$ & Gasteiger partial charge  \\
        \midrule \midrule
        \multicolumn{2}{c}{\textbf{representation L3~\cite{duvenaud2015convolutional}}} \\
        size & description \\
        \midrule
        $44$ & one-hot vector specifying the type of atom  \\
        $6$ & number of heavy neighbours as one-hot vector \\
        $5$ & number of hydrogen atoms as one-hot vector  \\
        $6$ & implicit valence as one-hot vector \\
        $1$ & is in aromatic system  \\
        \midrule \midrule
        \multicolumn{2}{c}{\textbf{representation L4~\cite{yang2019analyzing}}} \\
        size & description \\
        \midrule
        $100$ & one-hot vector specifying the type of atom  \\
        $6$ & number of bonds the atom is involved in as one-hot vector \\
        $5$ & formal charge as one-hot vector \\
        $4$ & chirality as one-hot vector \\
        $5$ & number of hydrogen atoms as one-hot vector  \\
        $5$ & hybridisation type as one-hot vector \\
        $1$ & is in aromatic system  \\
        $1$ & atomic mass \\
        \bottomrule
    \end{tabular}
    \label{tab:atom_emb_lit}
\end{table}

\begin{figure}[tbh!]
    \centering
    \includegraphics[width=.75\linewidth]{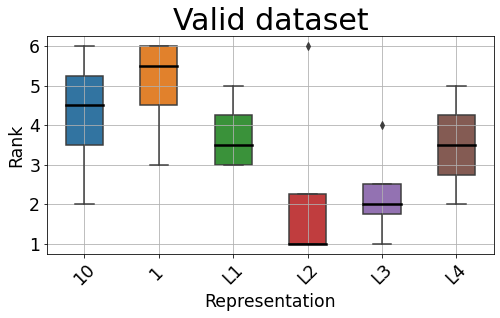}
    \includegraphics[width=.75\linewidth]{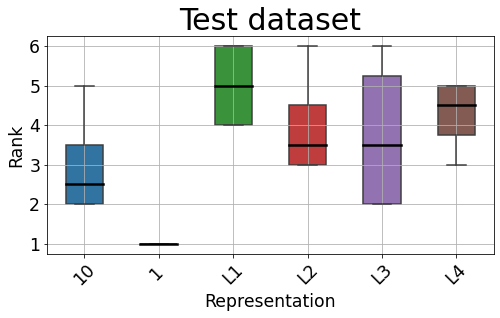}
\caption{Rankings obtained for each representation on all datasets for validation (left) and test (right) datasets. The median ranking is marked with a bold line.}
\label{fig:scores_lit_repr}
\end{figure}

For each representation, we report the validation and test results of the model with the best result on the validation set. The results are presented in Table~\ref{tab:scores_lit_repr} and rank-plots are depicted in Figure~\ref{fig:scores_lit_repr}. Additionally, we include results of representation 10 (no formal charge) which on average performed best in the previous section.
All experiments were conducted using the same settings as in the previous section.

\begin{table*}[htb!]
    \caption{Average valid and test mean squared error of models trained with different representations from literature. Best mean test results are in bold.}
    \centering
    \begin{tabular}{ c | cc | cc | cc | cc }
        \hline
        {} & \multicolumn{2}{c | }{Rat} & \multicolumn{2}{c |}{QM9} & \multicolumn{2}{c |}{ESOL (scaffold)} & \multicolumn{2}{c}{ESOL (random)} \\
        Representation & valid & test & valid & test & valid & test & valid & test \\
        \hline
        10                                 & $.254 \pm .02$ & $.247 \pm .03$ & $2.667 \pm .76$ & $9.698 \pm 1.47$ & $.110 \pm .01$ & $.201 \pm .01$ & $.085 \pm .01$ & $.123 \pm .01$ \\
        1~\cite{coley2017convolutional}    & $.226 \pm .02$ & $\mathbf{.214 \pm .01}$ & $4.531 \pm .19$ & $\mathbf{9.193 \pm 1.18}$ & $.107 \pm .01$ & $\mathbf{.166 \pm .02}$ & $.086 \pm .01$ & $\mathbf{.115 \pm .01}$ \\
        L1~\cite{liu2019chemi}              & $.221 \pm .02$ & $.286 \pm .02$ & $2.813 \pm .77$ & $20.544 \pm 6.85$ & $.111 \pm .01$ & $.238 \pm .02$ & $.078 \pm .01$ & $.133 \pm .01$ \\
        L2~\cite{li2019deepchemstable}      & $.202 \pm .01$ & $.239 \pm .02$ & $2.087 \pm .16$ & $17.52 \pm 4.74 $ & $.115 \pm .01$ & $.216 \pm .02$ & $.073 \pm .01$ & $.141 \pm .01$ \\
        L3~\cite{duvenaud2015convolutional} & $.204 \pm .02$ & $\mathbf{.214 \pm .02}$ & $3.308 \pm 1.19$ & $30.036 \pm 3.21$ & $.103 \pm .01$ & $.225 \pm .02$ & $.076 \pm .01$ & $.117 \pm .01$ \\
        L4~\cite{yang2019analyzing}         & $.207 \pm .04$ & $.224 \pm .02$ & $3.831 \pm 1.18$ & $25.967 \pm 3.54$ & $.105 \pm .01$ & $.223 \pm .02$ & $.078 \pm .01$ & $.134 \pm .01$ \\
        \hline
    \end{tabular}
    \label{tab:scores_lit_repr}
\end{table*}

Representation 1 obtained the best test scores on all of the datasets despite having the smallest size among the literature representations under analysis. More interestingly, it also consistently achieves a much lower generalisation gap than the remaining literature representations.
Representation 10, which differs only by dismissing information about formal charge, consistently achieves similar but poorer results, and the generalisation gap also slightly increases. We see this as an indication of good generalisation properties of this representation.

Representations L1 and L2 despite being similar in size to representation 1 (respectively, 33 and 38 versus 26) usually achieve weaker results and with a much higher generalisation gap. Therefore, we conclude that the generalisation properties cannot be attributed solely to the size of the representation.

Not surprisingly the results on ESOL with scaffold split are consistently lower than on ESOL random and with a higher generalisation gap which should be attributed to the method of the data split. 

Overall, we see that a single representation can consistently outperform others on multiple tasks though usually the differences are not strongly emphasised.

\section{Conclusions}

In this study, we examine the influence of atomic representations on the predictive performance of graph neural networks.
We show that the choice of atom features used in the representation results in an improved or reduced performance of the trained models and we confirm the significance of the arising differences by one-tailed Wilcoxon test. The differences are most pronounced in case of the QM9 dataset.

The presented results indicate that the choice of atom features is task-specific though some general conclusions can be drawn. Not surprisingly representations with more features tend to give better results. However, removing features related to aromaticity, inclusion in a ring, and formal charges can improve the model quality. On the other hand, adding information about heavy neighbours and hydrogens gives the biggest performance boost across all datasets.

The qualitative analysis suggests that the committed errors can be attributed to the absence of information about atom features which were not included in the model's representation.

Comparing representations used in literature reveals that a single representation can consistently outperform others though the differences are not strongly emphasised.
We also take a look at generalisation properties of selected representations and conclude that they cannot be attributed solely to the size of the representation.

To the best of our knowledge, this is the first methodological study that focuses on the relevance of atom representation to the predictive performance of graph neural networks.

\bibliographystyle{IEEEtran}
\bibliography{sample}

\end{document}